\documentclass{mem}
\usepackage{natbib}\usepackage{txfonts}\usepackage{balance}
\usepackage{graphicx}
\usepackage{subfigure}
\usepackage[a4paper]{hyperref}
\idline{75}{282}
\begin{document}
\def\teff{$T\rm_{eff }$}
\def\kms{$\mathrm {km s}^{-1}$}

\title{
Turbulent AGN tori
}

   \subtitle{}

\author{
M. \,Schartmann\inst{1,2,3} 
\and K. \, Meisenheimer\inst{1}
\and H. \, Klahr\inst{1}
\and M. \, Camenzind\inst{4}
\and S. \, Wolf\inst{5}
\and Th.\, Henning\inst{1}
         }

\offprints{M. Schartmann}

\institute{
Max-Planck-Institut f. Astronomie, K\"onigstuhl 17, D-69117 Heidelberg, Germany
\and Max-Planck-Institut f. extraterr. Physik, Giessenbachstra\ss e, D-85748 Garching, Germany
\and Universit\"ats-Sternwarte M\"unchen, Scheinerstra\ss e 1, D-81679 M\"unchen, Germany
\and ZAH, Landessternwarte Heidelberg, K\"onigstuhl 12, D-69117 Heidelberg, Germany
\and Christian-Albrechts-Universit\"at zu Kiel, Leibnizstr.15, D-24098 Kiel, Germany\\ 
\email{schartmann@mpe.mpg.de}
}

\authorrunning{Schartmann }

\titlerunning{Turbulent AGN tori}

\abstract{
Recently, the MID-infrared Interferometric instrument (MIDI) at the VLTI has shown that dust tori in 
the two nearby Seyfert galaxies NGC~1068 and the Circinus galaxy are geometrically thick and 
can be well described by a thin, warm central disk, surrounded by a colder and fluffy
torus component. By carrying out hydrodynamical simulations with the help of the TRAMP code 
\citep{Klahr_99}, we 
follow the evolution of a young nuclear star cluster in terms of discrete mass-loss and energy injection 
from stellar processes. This naturally leads to a filamentary large scale torus component, where 
cold gas is able to flow radially inwards. The filaments open out into a dense and very turbulent 
disk structure. In a post-processing step, we calculate observable quantities like spectral energy
distributions or images with the help of the 3D radiative transfer code MC3D \citep{Wolf_03}. 
Good agreement is found
in comparisons with data due to the existence of almost dust-free lines of sight through the large scale 
component and the large column densities caused by the dense disk.    
}
\maketitle{}

\section{Introduction}
Within the current model of Active Galactic Nuclei (AGN),
two observed classes of galaxies can be explained by one intrinsically unique 
AGN model, the so-called {\it Unified Scheme of Active
Galactic Nuclei}: a hot accretion disk surrounds the central black hole and crosses over into a geometrically 
thick torus. Nowadays, it is thought that the gap between these two is filled by the so-called Broad Line Region, 
see \citet{Gaskell_08}. The optically thick torus blocks the sight towards the centre, 
when viewed edge-on and enables the viewing of the characteristic emission peak of the accretion disk and 
the Broad Line Region clouds when viewed face-on.
Numerous observations back this model and also hint towards a clumpy nature of the gas and dust torus.

\section{Our hydrodynamical model}     

\begin{table*}[t!]
\begin{center}
\caption[Parameters of our standard model]{Parameters of our standard model.}
 \label{tab:stanpar}
\begin{tabular}{cc|cc|cc}
\hline
Parameter & Value & Parameter & Value & Parameter & Value \\
\hline
$M_{\mathrm{BH}}$ & $6.6\,\cdot 10^{7}\,M_{\sun}$ & $R_{\mathrm{T}}$ & 5\,pc & $T_{\mathrm{ini}}$ & $2.0\,\cdot 10^{6}\,$K \\
$M_{\mathrm{Rc}}$ & $6.7\,\cdot 10^{8}\,M_{\sun}$ & $\sigma_{*}$ & 165\,km/s  &  $\dot{M}_{\mathrm{n}}$ & $6.0\,\cdot 10^{-9}\,M_{\sun}/(yr\,M_{\sun})$\\
$M_{\mathrm{gas}}^{\mathrm{ini}}$ & $1.2\,\cdot 10^{4}\,M_{\sun}$ &
 $\alpha$ & 0.5 & $M_{\mathrm{PN}}$ & $2.2\,M_{\sun}$\\
$R_{\mathrm{c}}$ & 25\,pc & $\Gamma$ & $5/3$ & SNR & $10^{-10}\, \mathrm{SNe}/(\mathrm{yr}\,M_{\sun})$\\
\hline
\end{tabular}
\end{center}

\medskip
 Mass of the black hole 
 ($M_{\mathrm{BH}}$), stellar mass within the core radius of the distribution ($M_{\mathrm{Rc}}$), 
 initial gas mass ($M_{\mathrm{gas}}^{\mathrm{ini}}$), cluster core 
 radius ($R_{\mathrm{c}}$), torus radius ($R_{\mathrm{T}}$),
 stellar velocity dispersion ($\sigma_{*}$), exponent of the angular momentum distribution ($\alpha$),
 adiabatic exponent ($\Gamma$), 
 initial gas temperature ($T_{\mathrm{ini}}$), normalised mass injection rate ($\dot{M}_{\mathrm{n}}$), 
 mass of a single injection ($M_{\mathrm{PN}}$) and the
 supernova rate (SNR).
\end{table*}

As was recently shown by \citet{Davies_07} with the help of adaptive optics techniques, a large number 
of Seyfert galaxies possess young nuclear star clusters. We assume that they are built-up in a short-duration 
starburst and are interested in the effects of the evolution of their stars after the violent phase of 
supernova type~II explosions is already over, after approximately 40~Myrs. After this phase, energy input into 
the interstellar medium is dominated by the ejection of planetary nebulae and energy input is provided by
supernova type~Ia explosions. We model both of these processes by discrete input of mass and energy into a 
number of grid cells. The gas blobs resembling planetary nebulae additionally get a certain bulk velocity,
made up of the turbulent velocity of the {\it hot} stellar system and the rotational velocity of the stars.
For the rate of mass loss, we use the approximative calculations of \citet{Jungwiert_01}. The supernova rate
is difficult to determine and hence a more or less free parameter in our simulations, constrained by 
observations and a large parameter study. 
The third major ingredient of our simulations is optically thin cooling, where we employ an effective cooling 
curve calculated with the help of the CLOUDY code \citep{Ferland_93} by \citet{Plewa_95}.
The parameters of our standard model are summarised in Table~\ref{tab:stanpar}.

\section{Resulting structure}

\begin{figure*}[htb!]
\centering
\resizebox{0.70\hsize}{!}{\includegraphics[clip=true]{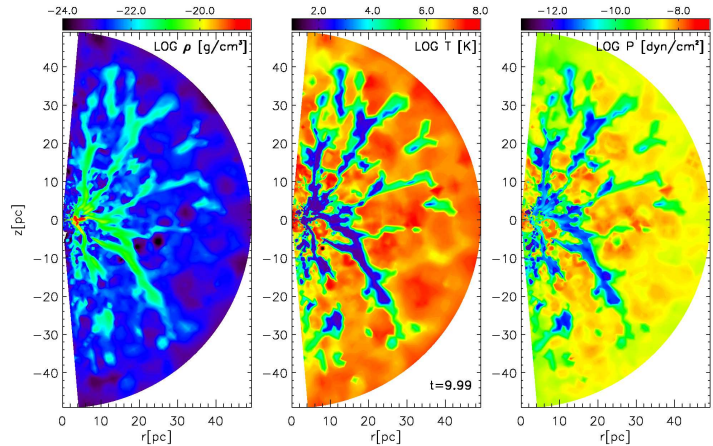}}
\caption[State of the torus after roughly ten global orbits, meridional plane]{\footnotesize State of the torus
    after roughly ten global orbits. Shown are from left to
    right the density distribution, the temperature distribution and the
    pressure distribution of a meridional slice through the 3D data cube. 
    All images are displayed in logarithmic scaling.}
\label{fig:fig03}
\end{figure*}
%

%
%

Within the first few percent of a global orbit of the simulation, single ejected planetary nebulae
are visible, as well as shock fronts due to recent supernova explosions. The turbulent nature of these 
processes, together with cooling instability soon leads to the formation of long filaments of cold gas
and voids of hot gas in-between, as evident from Fig.~\ref{fig:fig03}(a). Gas moves mainly radially inwards 
within the cold filaments, leading to a separation of gas velocities between different temperature 
phases. The largest infall velocities are reached by the cold material and the smallest by the hot 
gas between the filaments. This is caused by the larger support against gravity by thermal pressure in the
hot phase and the additional turbulent pressure, arising from the perpetual stirring by supernova explosions
and the input of mass with high random velocities. Therefore, cold gas moves radially inwards, until the 
angular momentum barrier is reached and a disk forms. Due to the infall of gas within the filaments from 
different directions, it shows turbulent behaviour. 
From Fig.~\ref{fig:fig03}(a,b) it is also evident that
density and temperature are complementary, which results from the implemented optically thin cooling, which
scales proportional to the square of the gas density. Pressure equilibrium is not reached, because of cooling 
instability in the filaments as well as recent supernova input, both acting on smaller timescale 
than relaxation processes (Fig.~\ref{fig:fig03}c). 
The resulting multi-phase medium reflects the shape of the cooling function as well as the local 
imbalance with energy and momentum input by supernovae and planetary nebulae.

\section{Comparison with observations}

\begin{figure*}[htb!]
\centering
\subfigure{\resizebox{0.45\linewidth}{!}{\includegraphics[clip=true]{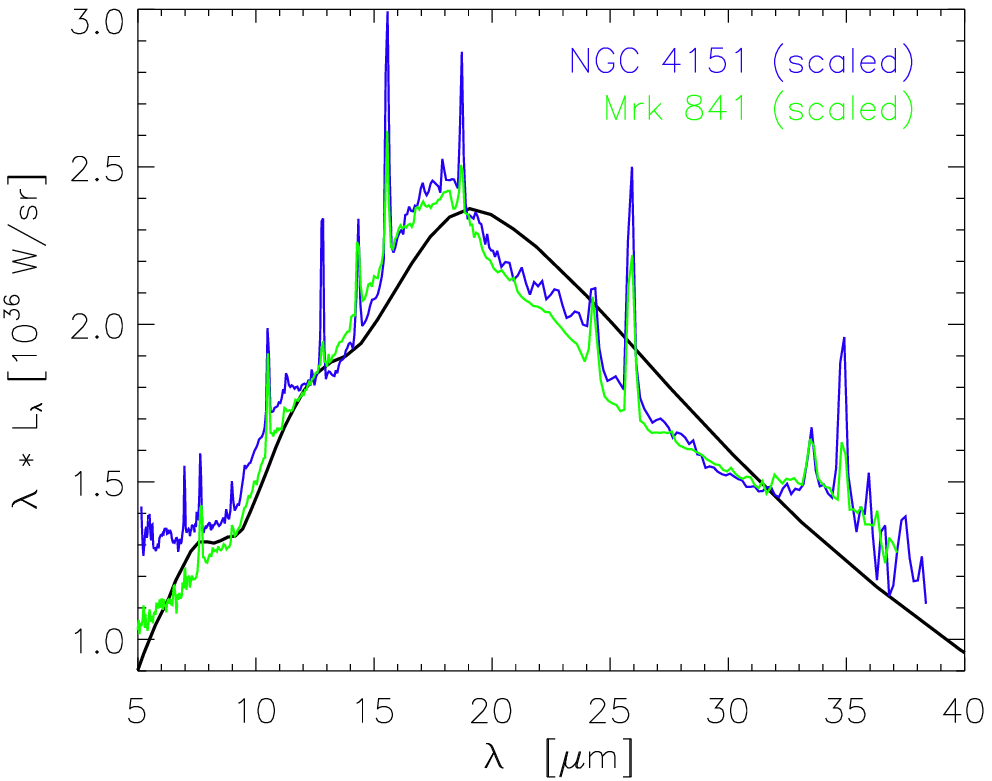}}
  \label{fig:fig10}
}
\subfigure{\resizebox{0.45\linewidth}{!}{\includegraphics[clip=true]{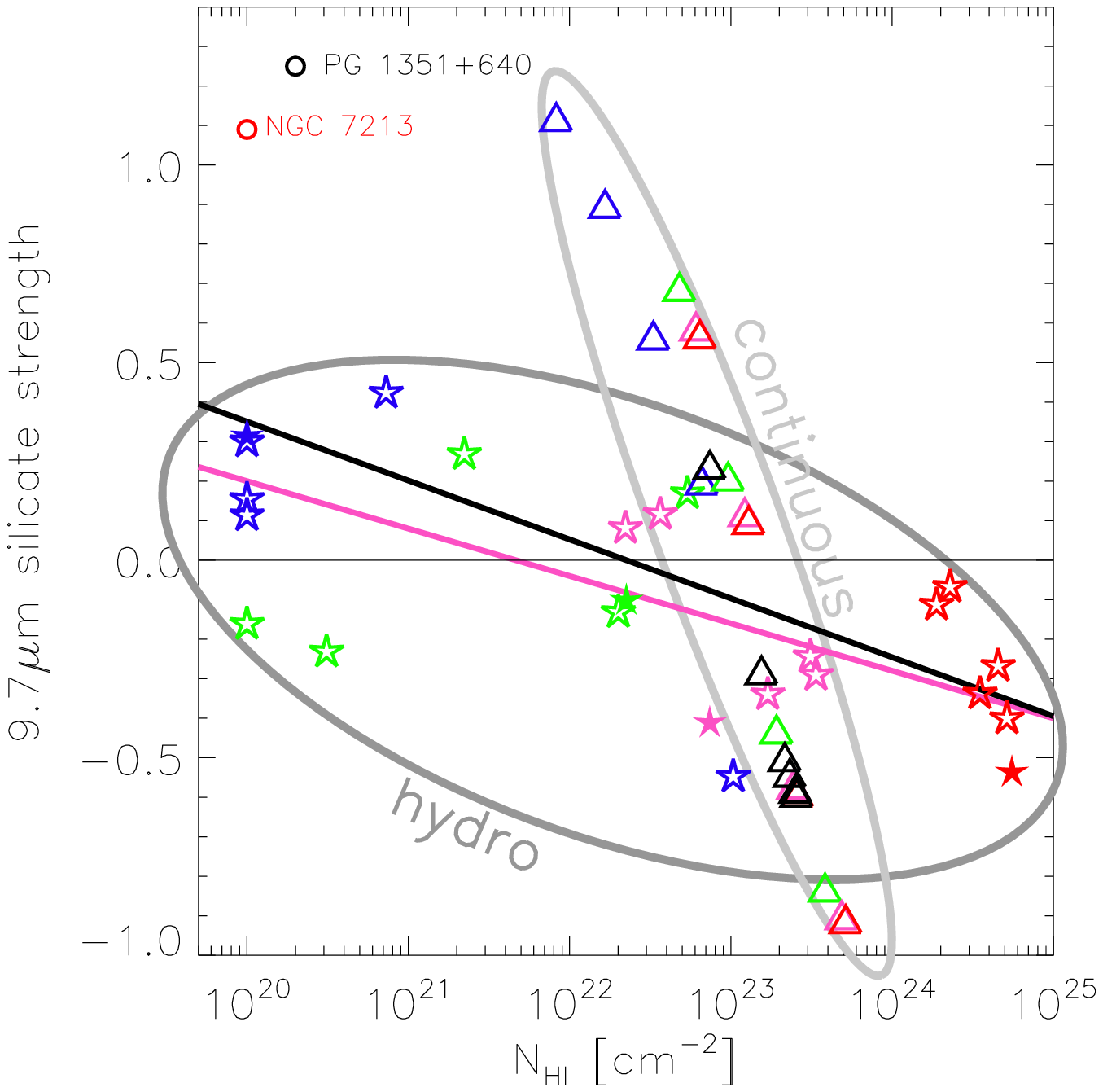}}
\label{fig:fig11}
}
\caption[Comparison with observations:]{
  {\bf a) } Comparison 
  of our standard model (solid line -- $i=30\degr$) with IRS {\it Spitzer} observations of
           NGC\,4151 \citep{Weedman_05} and Mrk\,841 (H.~W.~W.~Spoon, private communication).
  {\bf b) } Comparison of resulting silicate feature strengths and column
           densities of our hydromodels (standard model plus supernova and mass loss rate study)
           given as stars and the continuous TTM-models (standard model
           and models from a dust mass study, see \citealp{Schartmann_05}) given as
           triangles. The filled stars denote our hydrodynamical standard
           model. Color denotes inclination angle (blue -- $8\degr$, green
           -- $30\degr$, magenta -- $60\degr$, red -- $90\degr$, black -- intermediate angles of the
           continuous standard model). 
           The solid lines
           are linear fits to the observational data shown in
           \citealp{Shi_06} (black -- all objects, magenta -- Seyfert
           galaxy sample).
}
\end{figure*}

With the help of a constant dust-to-gas ratio, we convert the resulting gas density distribution 
(see Fig.~\ref{fig:fig03}a) into the corresponding dust density distribution. The latter 
is used in radiative transfer calculations with MC3D to calculate observable quantities like spectral 
energy distributions or images at various wavelengths. Fig.~\ref{fig:fig10} shows the comparison of 
our standard model at $30\degr$ inclination angle with two intermediate type Seyfert galaxies (scaled)
observed with the Spitzer satellite. The largest deviations at around $35\mu$m are caused by crystalline
forsterite in emission, which is currently not included in our dust model. 
As was shown by \citet{Shi_06}, AGN follow a linear relation between neutral hydrogen column density 
and the $10\,\mu$m silicate feature strength. We determine the latter as the difference between the 
simulated spectral energy distribution (SED) and the underlying (spline-fitted) continuum. 
Fig.~\ref{fig:fig11} shows the comparison of the observed relation for the whole \citet{Shi_06} sample 
(black line) and the Seyfert galaxies alone (magenta) and the relation resulting from our 
hydrodynamical simulations described above. We obtain a shallow slope and scatter comparable to 
the observations. In contrast to this, our previous continuous torus models as described in 
\citet{Schartmann_05} are not able to explain the full range of neutral hydrogen column densities
without resulting in unphysically silicate feature strengths. Additionally, continuous models are  
in disagreement with the observed large scatter of the relation. Therefore, it can be taken 
as a further observational indication for clumpiness of AGN tori. The reason for the 
good comparison of our hydrodynamical models is caused by the two component structure, made up of 
a dense thin disk and a fluffy large scale torus. The filamentary structure guarantees 
the existence of more or less unobscured lines of sight towards the funnel region, where the silicate feature in 
emission is produced and thereby relativises the silicate feature in absorption. The smearing out 
of the silicate feature in emission due to clumpy dust distributions is already a well known issue 
(see e.~g.~\citealp{Nenkova_02}, \citealp{Hoenig_06} or \citealp{Schartmann_08}). 
The dense disk component enables also large
hydrogen column densities. 

%

%

\section{Summary}

A physical model for the built-up and evolution of gas tori in Seyfert galaxies is
presented. We follow the evolution of a young nuclear star cluster, during the 
phase, where the ejection of planetary nebulae is the dominant process for mass input. 
Together with supernova energy input and the onset of cooling instability yield
a large scale, filamentary torus component with cavities
of hot gas in-between. Within the filaments, gas cools and gets accreted in radial 
direction. Close to the angular momentum barrier, it opens out into a dense disk 
component with turbulent character due to the inflow of filaments and clumps of gas 
from random directions. This two component structure enables a good comparison with 
observations.

\begin{acknowledgements}
We are grateful to H.W.W.Spoon for providing us the reduced {\it Spitzer} SED of Mrk~841 and 
K.R.W.Tristram for helpful discussion. 
\end{acknowledgements}

\bibliographystyle{aa}
\bibliography{astrings,literature}

\end{document}